\documentclass[useAMS,usenatbib]{mn2e}

\usepackage{multirow}
\usepackage{multicol}
\usepackage{graphicx}

\title[A black hole in XTE\,J1859+226]{Evidence for a black hole in the X-ray
transient XTE\,J1859+226}
\author[Corral-Santana et al.]
{J. M. Corral-Santana$^{1,2}$
\thanks{E-mail: jcorral@iac.es (JMC-S), jcv@iac.es (JC), tsh@iac.es (TS), czurita@iac.es (CZ), igm@iac.es (IGM-P), prguez@iac.es (PR-G)},
J. Casares$^{1,2}$, T. Shahbaz$^{1,2}$,
\newauthor
C. Zurita$^{1,2}$, I. G. Mart\'inez-Pais$^{1,2}$, P. Rodr\'iguez-Gil$^{3,1,2}$
\\
\\
$^{1}$Instituto de Astrof\'isica de Canarias (IAC), V\'ia L\'actea s/n, La Laguna E-38205, S/C de Tenerife, Spain\\
$^{2}$Departamento de Astrof\'isica, Universidad de La Laguna, La Laguna E-38205, S/C de Tenerife, Spain\\
$^{3}$Isaac Newton Group of Telescopes, Apartado de Correos 321, Santa Cruz de La Palma E-38700, Spain\\
}

\begin{document}

\date{Accepted 2011 January 31.}

\pagerange{\pageref{firstpage}--\pageref{lastpage}} \pubyear{2011}

\maketitle

\label{firstpage}

\hyphenation{pro-mi-sing mi-ni-ma in-di-vi-dual co-rres-pond 
co-ve-red spec-tra pre-pa-red du-ring e-ve-ry con-si-de-ra-ble 
Fi-lip-pen-ko a-na-ly-sis ve-lo-ci-ty fli-cke-ring pre-sen-ce 
ack-now-led-ge co-ve-ring ca-li-bra-tion res-trict pho-to-me-try 
pa-ra-me-ters se-con-da-ry com-bi-ning du-ring chan-ging si-mu-la-ted
re-a-lis-tic ins-ti-tu-to spec-tros-co-pic dy-na-mi-cal a-ve-ra-ge con-ti-nu-um
sub-trac-ting do-mi-na-ted mi-nis-try un-ders-tan-ding knowl-edge com-pa-nion 
re-fe-reed pho-to-me-tric res-pec-tive-ly flick-er-ing qui-es-cent li-mi-ted
pa-ra-me-ter re-a-lis-tic mi-ni-mum u-sing ins-tal-led}

\begin{abstract}
We present the results of time-resolved optical photometry and spectroscopy
of the X-ray  transient XTE\,J1859+226 (V406\,Vul). Photometric
observations taken during  2000 and 2008 reveals the presence of the secondary
star's ellipsoidal
modulation. Further photometry obtained in 2010
shows the system $\simeq$1\,mag brighter  than its quiescence level and the
ellipsoidal modulation diluted by strong flaring activity. Spectroscopic
data obtained with the 10.4-m GTC in 2010 reveals  radial velocity variations of
$\sim 500\,\rm km\,s^{-1}$ over 3\,h. A simultaneous fit to the photometry and
spectroscopy using sinusoids to represent the secondary star's ellipsoidal and
radial velocity variations, yields an orbital period of 
$6.58\pm 0.05\,\rm h$ and a secondary
star's radial velocity semi-amplitude of $K_2= 541 \pm 70~\rm km\,s^{-1}$.
The implied mass function is $f(M)=4.5 \pm 0.6~\rm M_\odot$, significantly
lower than previously reported but consistent with the presence of a black
hole in XTE\,J1859+226. The lack of eclipses sets an upper limit to the inclination of 70 degrees
which yields a lower limit to the black hole mass of $5.42\,\rm M_\odot$.
\end{abstract}


\section{INTRODUCTION}
\label{sec:intro}
The mass distribution of compact objects has crucial impact on fundamental
physics, and can only be determined from the study of X-ray binaries \citep{Charles2006,Casares2007}. 
In particular, the theoretical black-hole (BH) and neutron
star (NS) mass distributions depend critically on the equation of state (which establishes 
the critical mass dividing NS and BH formation) and on our current understanding 
of the late stages in the evolution of massive stars and supernovae. The latter
contains several theoretical uncertainties which completely dominate the final mass
distribution 
\citep[e.g. treatment of convective mixing, amount of fall-back, mass-cut, mass-loss 
during Wolf-Rayet phase, initial mass function of progenitors, etc; see e.g.][]{Fryer2001}. 
It should be noted that the upper mass cut-off of the BH distribution
is particularly constraining \citep{Orosz2007}.

Only 16 systems out of an estimated Galactic population of $\sim10^{3}$ 
transient X-ray binaries have reliable dynamical mass determinations (e.g. \citealt{Casares2010}).  
The typical error bars in BH mass measurements are $\sim$40\,per cent and it is clear that more BH 
discoveries and a factor 2-3 improvement is needed before any constraints can be set on 
supernovae models \citep{Bailyn1998,Ozel2010}.
The mass determination in X-ray binaries requires knowledge of the radial velocity
curve, the binary mass ratio and the orbital inclination (the latter from modelling the tidally distorted 
companion's light curve).

XTE J1859+226 was discovered during its 1999 outburst \citep{Wood1999}
and its X-ray properties promptly classified it as a BH candidate. Orbital periods of
6.72\,h \citep{Uemura1999}, 9.12\,h \citep{Garnavich1999} and 18.72\,h \citep{McClintock2000a} 
were reported although none of them could be confirmed. Finally, data
taken in quiescence \citep{Zurita2002} during September and November 2000 (in 4 night
snapshots of 1.5\,h-4\,h duration) exhibited a $\sim$0.2\,mag semi-amplitude sinusoidal
modulation, consistent with an secondary star's ellipsoidal variation (caused by the changing
visibility of the tidally distorted companion star as it orbits the compact object). The
periodogram of this modulation suggests that periods from 
6.6 to 11.2\,h are equally
possible at the 68\,per cent confidence level. On the other hand, \cite{Filippenko2001} 
report a preliminary analysis of radial velocities based on 10 spectra obtained
over two nights. They claim evidence for a sinusoidal variation with a period of   
9.1\,h and companion radial velocity semi-amplitude $K_2 = 570 \pm 27$\,km\,s$^{-1}$.
The implied mass function is $f (M) = 7.4 \pm 1.1~\rm{M_\odot}$,
one of the largest ever measured making J1859+226 a very promising massive BH. 
Although this result has been used by other authors, it should be noted that it
has never been published in a refereed journal and hence we believe these 
parameters require confirmation. Therefore, we have
embarked on a photometric and spectroscopic campaign with the aim of 
determining the orbital period of the binary and its dynamical mass function.

\section{OBSERVATIONS AND DATA REDUCTION}
\label{sec:obs}
\subsection{Photometry}
\label{subsec:obsphotom}
XTE\,J1859+226 was observed on 2008 July 31-August 1 and 2010 July 13-14 using 
ALFOSC at the 2.5-m Nordic Optical Telescope (NOT) in the $R$-band. 
The conditions for the 2008 campaign were affected by thick dust (calima) 
and therefore we
decided to use long integration times. Eight 1800-s images were obtained on the
night of July 31 and thirteen 900-s images on August 1, covering $\sim$5-7\,h per night.  
The 2010 campaign was performed under excellent seeing ($\sim$0.5\,arcsec) and
transparency conditions, so shorter exposure times were adopted. Twenty
two and thirty two 800-s integrations were obtained on the nights of July 13-14
respectively, covering $\sim$8\,h per night. 
XTE\,J1859+226 was also observed on the night of 2010 August 8 using ACAM at the 
4.2-m William Herschel Telescope (WHT). A total of 147 frames were obtained in 
the $R$-band using 120-s integration time for $\sim$6.5\,h. 
All the images were corrected for bias and flat-fielded with \verb+IRAF+ in the 
standard way. Aperture photometry was then performed on the object and the five 
comparison stars reported in \cite{Zurita2002}.

\subsection{Spectroscopy}
\label{subsec:obsspec}
Optical spectroscopy of XTE\,J1859+226 was obtained on the nights of 2010  July
17 and August 13 using OSIRIS at the 10.4-m Gran Telescopio Canarias  (GTC).
Five 1900-s spectra were obtained every   night using the R1000B grism and a
0.6-arcsec slit, which provides a  wavelength coverage of $\lambda
\lambda$3630-7500 at $255$\,km\,s$^{-1}$  resolution (FWHM).  The radial velocity
standards 61 Cyg A \& B were also observed during the  second run using an
identical set up. A Hg-Ar-Ne arc was obtained with the telescope in park
position to provide the  wavelength calibration scale.  Standard procedures were
used to de-bias and flat-field the spectra. The 1-dimensional spectra were
extracted using optimal extraction routines which maximize the final
signal-to-noise ratio. The wavelength calibration, flux calibration and radial
velocity analysis were made with 
\verb+MOLLY+\footnote{http://www.warwick.ac.uk/go/trmarsh/software}.  The
wavelength calibration was obtained using a fourth-order polynomial fit to 34
arc lines resulting in a dispersion of 2.14\,\AA~pix$^{-1}$ and a rms scatter
of 0.07\,\AA. The instrumental flexure was monitored by measuring the position
of the sky line  OI\,5577.340\AA~ which was used  to correct the individual
spectra.  The spectra were also calibrated in flux using observations of the
flux standard GL57-34. 
%
\begin{figure}
\centering
	\begin{center}
		\includegraphics[scale=0.45]{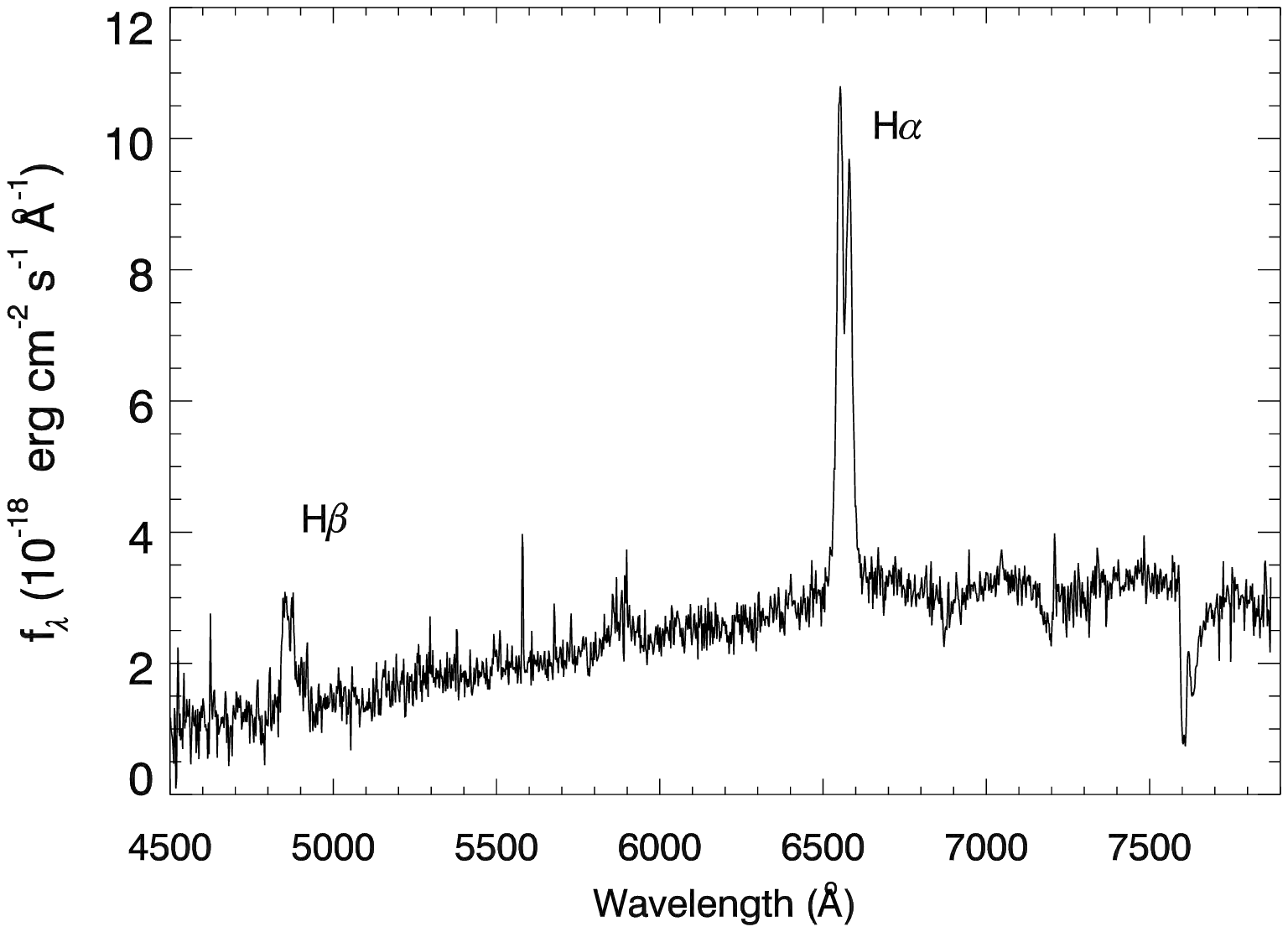}
		\includegraphics[scale=0.45]{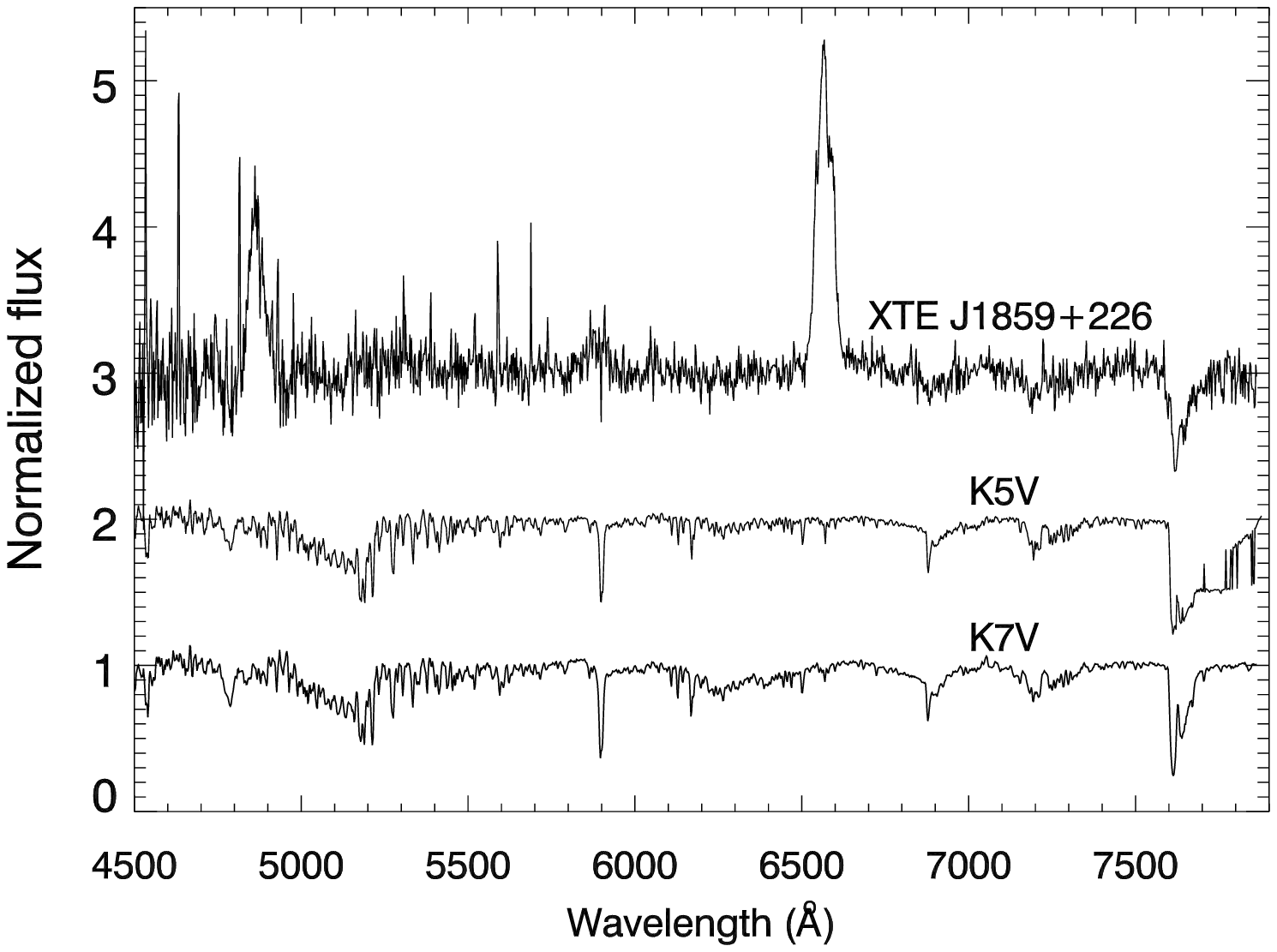}
		\caption{{\bf Top:} the average optical spectrum of XTE\,J1859+226. 
		A red continuum with broad double-peaked emission lines 
		of H$\alpha$ and H$\beta$ are clearly present. 
		The feature at $\sim$5890\,\AA~is possibly associated to HeI 
		emission. {\bf Bottom:} Comparison of the Doppler corrected
		average of J1859+226 with the two observed templates. The
		spectra have been rectified to the continuum and offset
		vertically for the sake of clarity.}
		\label{average}
	\end{center}
\end{figure}
\begin{figure}
\centering
	\begin{center}
		\includegraphics[scale=0.46]{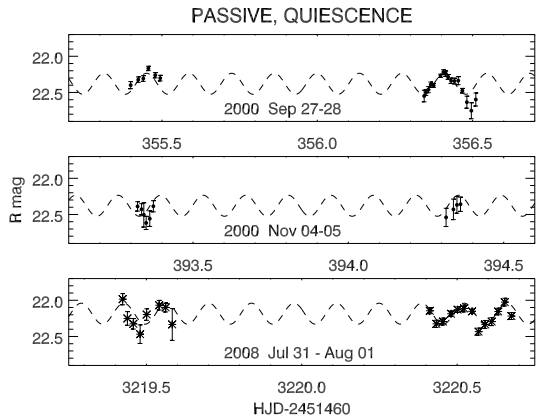}
		\caption{\textbf{\textit{Top \& middle panels:}} Zurita et al. (2002) epoch C optical photometry taken in 2000.
		\textbf{\textit{Bottom:}} Optical photometry taken in 2008 July 31 and 
		August 1 showing the dominance of the ellipsoidal modulation. The
		solid line is the simulated secondary star's ellipsoidal 
		modulation determined in section~\ref{sec:parameters}.
                Passive refers to the "true" quiescence state as defined by Cantrell \& Bailyn (2007).}
		\label{quiescence}
	\end{center}
\end{figure}

\section{OVERVIEW OF OPTICAL PHOTOMETRY AND SPECTROSCOPY}
\label{sec:photom}

\subsection{Spectroscopy}
\label{subsec:overspec}

The top panel in Fig.~\ref{average} shows the flux-calibrated average spectrum of  
XTE\,J1859+226. It shows a red continuum with broad double-peaked emission lines
at the positions of H$\alpha$ and H$\beta$, characteristic of  X-ray transients
in quiescence. The H$\alpha$ line has a typical equivalent width of
$\simeq$130\,\AA\, and FWHM$\simeq$2340\,km\,s$^{-1}$. The individual spectra
of XTE\,J1859+226 and the template stars 61\,Cyg\,A and B  were prepared for the
cross-correlation analysis by subtracting of a low-order  spline fit to the
continuum, after masking the H$\alpha$ and H$\beta$  emission lines and
atmospheric absorption features. The spectra were then rebinned onto  a uniform velocity
scale of 112\,km\,s$^{-1}$ per pixel and the unit continuum  was subtracted. 
Cross-correlation between the target spectrum and the K5V template 61 Cyg A was
then performed in the region 5000-6516\,\AA~after masking the atmospheric and
interstellar feature at $\lambda\lambda$6285-6310 and the bump at
$\sim$5890\,\AA, possibly associated to HeI emission. Radial velocities were
extracted by fitting parabolic functions to the cross-correlation peaks. Clear
velocity excursions of a few hundred km\,s$^{-1}$ are seen on each night. 
We note that the radial velocities are not significantly affected by the
choice of either a K5V or K7V template. However, in an attempt to constrain the
spectral type we have performed an optimal subtraction analysis of the Doppler 
corrected average of J1859+226 (using the orbital solution reported in
section~\ref{sec:parameters}) with the 2 templates (see e.g. \citealt*{Marsh1994}). The 
$\chi^2$ of the residual slightly favours the K5V template, with 928 versus 
944 for 742 degrees of freedom. The bottom panel of Fig.~\ref{average}
displays the Doppler corrected average and the 2 templates. Although the
spectrum is too noisy to make a direct comparison of the metallic absorption
lines, the relative depth of the broad CaH and TiO bands at 6300, 6800 and 
7200\,\AA~also supports the K5V versus the K7V spectrum.

\subsection{Photometry}
\label{subsec:overphotom}

\begin{figure}
\centering
	\begin{center}
		\includegraphics[scale=0.46]{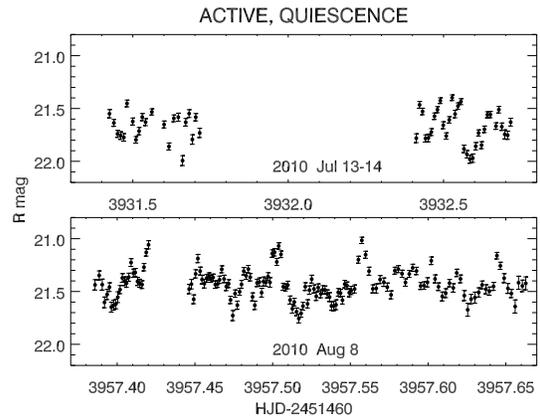}
		\caption{The optical photometry of XTE\,J1859+226 taken on 2010 
		July 13-14 using the NOT (top) and August 8 using the WHT (bottom). 
		The short 120-s integration time of the WHT images enables us to
		resolve the flickering activity which strongly dilutes the 
		secondary star's ellipsoidal modulation. Active refers to the
		flickering state as defined by Cantrell \& Bailyn (2007).}
		\label{flickering}
	\end{center}
\end{figure}

The light curves of XTE\,J1859+226 were determined through differential photometry
using 5 comparison stars (see \citealt{Zurita2002}) as local standards. 
\cite{Zurita2002} report a quiescent magnitude R=22.48$\pm$0.07 but we note
that the target is brighter in our photometric campaigns by $\sim$0.25
mag in 2008 and $\sim$0.8-1.0\,mag in 2010. The 2008 data clearly shows an 
ellipsoidal modulation (Fig.~\ref{quiescence}), caused by the tidal and rotational distortion of the
companion star (e.g. see Cen X-4 in \citealt{Shahbaz1993}). 
However, the  2010 light curves show 
considerable scatter due to the presence of strong aperiodic variability which
distorts the underlying ellipsoidal modulation: 
the flickering activity is most prominent in the 2010 August data (Fig.~\ref{flickering}, bottom panel)
compared with the 2010 July campaign (Fig.~\ref{flickering}, top panel). 
This is probably due to the shorter integration times 
which enables us to 
resolve the individual flare events. It is also when XTE\,J1859+226 is 
brighter, about 1\,mag above quiescence, so it is possible that the level of 
aperiodic variability correlates with the target brightness.
A similar behaviour is seen in the black hole binary A0620-00, where 
flickering is found to decrease with magnitude 
\citep{Cantrell2008, Cantrell2010}. 

The light curves of quiescent X-ray binaries are known to be contaminated by optical
flickering with time-scale of minutes to hours and amplitudes in the range 0.06-0.6\,mag 
(\citealt*{Zurita2003}, \citealt{Hynes2003, Shahbaz2005, Shahbaz2010}). For
comparison, the flares in XTE\,J1859+226 have a characteristic time-scale of
$\sim$10\,min and amplitudes up to 0.5\,mag.
Therefore, for the sake of constraining the orbital period (see section~\ref{sec:parameters}) we 
decided to restrict our analysis to the 2008 campaign, when the target was closer 
to its quiescent level and showed minimum flickering activity (the \textit{passive} state as defined by
\citealt{Cantrell2007}). 
These data were also combined with 
quiescent photometry obtained in 2000 and reported in \cite{Zurita2002}.
Both datasets were detrended by subtracting the nightly mean value.\\

\section{PERIOD ANALYSIS AND DETERMINATION OF ORBITAL PARAMETERS}
\label{sec:parameters}
To determine the orbital period of XTE J1859+226 we compute a $\chi^2$ 
periodogram using a prior knowledge about the expected modulations observed in
the binary system. 
For a fixed frequency, we simultaneously fit the photometric
light curve 
(32 and 21 data points from the 2000 and 2008 campaigns 
respectively; $N_{LC}$) and the
radial velocity curve (10 data points; $N_{RV}$) with a model to represent the 
photometric and radial velocity variations. The photometric ellipsoidal 
modulation is equivalent to sinusoid of frequency 2$f$ (where $f$ is the orbital
frequency) which lags by 90 degrees in phase relative to a sinusoid at frequency $f$. 
Although the light curves were taken when the system was in the "passive 
quiescent" state, there is the possibility that the accretion disc contribution was not the
same for the two epochs, and so the light curves will have different amplitudes.
To account for the different amplitudes we first fit the light curve of 
\cite{Zurita2002} and then fit a scaled version of this to the data taken in our 2008 campaign.
The radial velocity motions of the secondary star can be modelled with a
sinusoid at frequency $f$. The phasing of the light curve and radial velocity
curve are related and so the free model parameters are the orbital frequency,
phase offset, the semi-amplitudes of the three sinusoids, 
the scale factor to
account for the different disc contamination between the light curves, 
the magnitude offset for the light curve and the systemic velocity for the radial 
velocity curve.

Although the Nyquist frequency sets the maximum frequency we can theoretically 
search, inspection of the
radial velocity curve shows a smooth modulation lasting more than 2\,h and
so any period we expect to detect should be longer  than this. Given that
observations are taken throughout entire and consecutive nights, 
the minimum frequency is limited to 24\,h.
Hence we limit our frequency search in the range 2 to 6\,cycles\,$\rm d^{-1}$ with a total of 126
frequency steps (2$\times N_{T}$; where $N_{T}$=$N_{LC}$+$N_{RV}$=63 is the total number of data
points; \citealt{Press2002}).
Given that there are two different types of data with different number of data
points, to optimise the fitting procedure we assigned relative weights to the
different data sets. After our initial search of the parameter space, which resulted
in a good solution, we scaled the uncertainties on each data set (i.e., the
light curve and the radial velocity curve) so that the total
reduced $\chi^2$ of the fit was $\approx$1 for each data set separately. After
the scaling, the fitting procedure were run again to produce the final set of
parameters.
The 99 per cent white noise significance levels was estimated using Monte Carlo
simulations. We generated light curves and radial velocities with exactly the
same sampling and integration times as the real data, added Gaussian noise using
the errors on the data points. We computed 5000 simulated light curves and then
calculated the 99 per cent confidence level at each frequency taking into 
account a realistic number of independent trial \citep{Vaughan2005}.  

\begin{figure*}
\centering
	\begin{center}
		\includegraphics[width=7.cm,height=16.5cm,angle=-90]{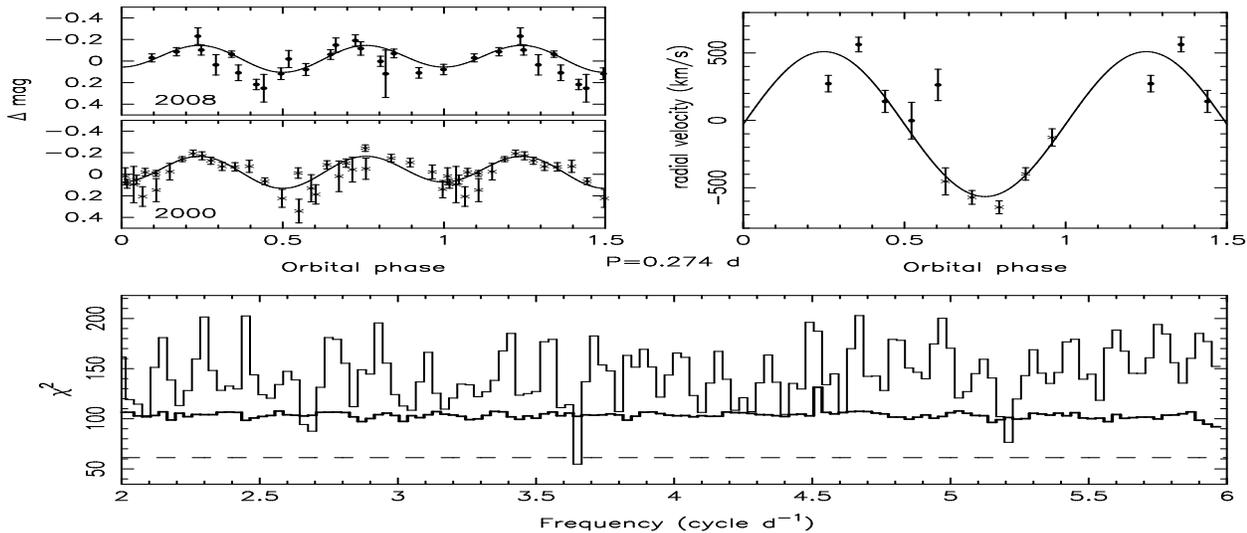}
		\caption{\textit{\textbf{Bottom:}} The $\chi^2$ periodogram. The solid line represents the 99 per cent 
		white noise significance levels. The dashed line shows the 99 per cent confidence level above the 
		minimum $\chi^2$ at a frequency of 3.65\,cycles\,$\rm d^{-1}$ (=0.274\,d). 
		\textit{\textbf{Top left and right:}} Respectively, the optical photometric 
		light curve and radial velocity curve phase folded on the 
		best fit period. 1.5 orbital cycles are shown for clarity. In the radial velocity curve
		plot, the filled circles and stars show the data taken on different nights. In the top
		light curve plot the filled circles are the data from 2008,
		whereas in the bottom light curve plot the 
		stars are the data from Zurita et al. (2002). 
		The solid line in the plots 
		show the best fit model using a double and 
		single sinusoid to simulate the secondary 
		star's ellipsoidal and radial velocity 
		variations (see section~\ref{sec:parameters}).}
         	\label{periodogram}
	\end{center}
\end{figure*}

Three minima above the noise level are clearly seen at 3.65, 5.21 and 
2.69\,cycles\,$\rm d^{-1}$ (see Fig.~\ref{periodogram}, bottom panel). However, the
best fit has a  $\chi^2$ of 54.7 (with 56 degrees of freedom) and occurs at a
frequency of 3.65\,cycles\,$\rm d^{-1}$.
The best fitted radial velocity parameters for this period are listed in Table~\ref{fitted}. Note
that the $K_2$ velocity, combined with the orbital period, implies a 
mass function $f(M)=K_2^3\,P / 2\pi G=4.5 \pm 0.6\,\rm M_\odot$ and hence
the dynamical confirmation of a black hole.
The next best fit at 5.21\,cycles\,d$^{-1}$ is only significant at the $3\times
10^{-4}$ per cent level  and the remaining peak at 2.69\,cycles\,d$^{-1}$ (i.e.
the $\sim 9.1\,\rm h$ period  reported by \citealt{Filippenko2001}) has an even
lower significance. The top panel in Fig.~\ref{periodogram} shows the photometric
light curve and radial velocity curve, folded on the 0.274\,d period and with the
best fitted solution superimposed.
The peak-to-peak amplitude of the light curve of \citep{Zurita2002} is
0.239\,mag and the difference between minima is 0.058\,mag. The scale factor
between the amplitudes of the two light curves is 0.84$\pm$0.20 which implies that
within the uncertainties, the two light curves have similar amplitudes and disc contribution.

The relatively large amplitude of the light curve and the possible 
difference between the
minima (where the minimum at phase 0.5 is deeper than the minimum at
 phase 0.0) places strong constraints on the binary inclination angle.
Using our X-ray binary model which predicts the light curve arising from a
Roche-lobe filling in an X-ray binary \citep{Shahbaz2003} we find that
systems with inclination angles of $\sim$40 degrees have equal minima.
Furthermore, the lack of eclipses gives an upper limit of $\sim$70 degrees,
which implies a lower limit in the mass of the black hole of $5.42\,\rm M_\odot$.
Our data are not sufficiently accurate to allow a detailed analysis of the
inclination angle, but we
can use the X-ray binary model to estimate the distance and inclination angle.
We assume a secondary star with an effective
temperature similar to that in A0620-00 (an X-ray transient with a 
K4V secondary star with an orbital
period of 7.8\,h; see \citealt*{Marsh1994}), 
E(B-V)=0.58\,\citep{Hynes2002} and the
orbital parameters derived in section~\ref{sec:photom}.
We find that to reproduce the large amplitude of the flux level of the
observed light curve requires a system at 14\,kpc with 
an inclination angle of 60 degrees and no accretion disc contribution, or a
system with an inclination angle of 70 degrees but with a disc contribution of
28\%. We note in passing that the Galactic rotation velocity at this 
distance is $\simeq -53\,\rm km\,s^{-1}$~\citep{Clemens1985} which is consistent with the observed systemic velocity
listed in Table~\ref{fitted}. Clearly a more accurate light curve is required before we can
place tight constraints on the model parameters.
%
\begin{table}
  \begin{tabular}{ccccc}
            $P$       &     $K_2$    &          $T_0$      &   $\gamma_0$\\
            (d)       &     ($\rm km\,s^{-1}$)   &      (HJD-2455000)  &   ($\rm km\,s^{-1}$)    \\
    \hline
    $0.274 \pm 0.002$ & $541 \pm 70$ & $395.689 \pm 0.002$ & $-28 \pm 53$ \\
  \end{tabular}
  \caption{List of the best fitted radial velocity parameters obtained for XTE J1859+226}
  \label{fitted}
\end{table}
%
\section{CONCLUSION}
\label{sec:conclusion}

We have presented $R$-band photometry of XTE\,J1859+226 obtained in 2008 and 
2010. The 2008 light curves are dominated by the classical ellipsoidal modulation.
However, the target is $\simeq$0.6--1\,mag brighter than its quiescent level
during 2010 and the secondary star's ellipsoidal modulation is found to be 
strongly diluted by the flickering activity. A $\chi^2$ periodogram analysis 
combining the radial velocities and the quiescent photometry yields an orbital period 
of $6.58\pm 0.05\,\rm h$  (lower than previous claims by \citealt{Filippenko2001})
and a velocity semi-amplitude  $541 \pm 70\,\rm km\,s^{-1}$. The implied mass
function is $4.5 \pm 0.6\,\rm M_\odot$ substantially lower than reported in
\cite{Filippenko2001}, but still consistent with the presence of a black hole. 
More observations are required to further refine the orbital parameters of 
XTE\,J1859+226, in particular, the mass function.

\vspace{-0.15cm}
\section*{Acknowledgements}
The use of Tom Marsh's \verb+MOLLY+ package is gratefully acknowledge.
The WHT is operated by the Isaac Newton Group (ING). 
The NOT is operated jointly by Denmark, Finland, Iceland, Norway, and Sweden.
Both telescopes together with the Gran Telescopio Canarias (GTC) are installed 
in the Spanish Observatorio del Roque de los Muchachos of the Instituto de 
Astrof\'isica de Canarias (IAC), in the island of La Palma.
We acknowledge support from the Spanish Ministry of 
Science and Innovation (MICINN) under the grant AYA\,2007--66887.
This program is also partially funded by the Spanish MICINN under the 
Consolider-Ingenio 2010 Program grant CSD2006-00070: "First Science with the GTC  (http://www.iac.es/consolider-ingenio-gtc)".
Based on observations made with the WHT Telescope on 2010 August 8 under the 
Spanish Instituto de Astrof\'isica de Canarias Director's Discretionary Time

\vspace{-0.35cm}

\bibliographystyle{mn2e}

\def\jnl@style{\it}                       
\def\mnref@jnl#1{{\jnl@style#1}}          

\def\aj{\mnref@jnl{AJ}}                   
\def\apj{\mnref@jnl{ApJ}}                 
\def\apjl{\mnref@jnl{ApJL}}               
\def\aap{\mnref@jnl{A\&A}}                
\def\mnras{\mnref@jnl{MNRAS}}             
\def\nat{\mnref@jnl{Nat.}}                
\def\iaucirc{\mnref@jnl{IAU~Circ.}}       
\def\atel{\mnref@jnl{ATel}}               
\def\iausymp{\mnref@jnl{IAU~Symp.}}       

\bibliography{j1859_jcorral_2011}

\begin{thebibliography}{}

\bibitem[\protect\citeauthoryear{{Bailyn}, {Jain}, {Coppi} \& {Orosz}}{{Bailyn}
  et~al.}{1998}]{Bailyn1998}
{Bailyn} C.~D.,  {Jain} R.~K.,  {Coppi} P.,    {Orosz} J.~A.,  1998, \apj, 499,
  367

\bibitem[\protect\citeauthoryear{{Cantrell} \& {Bailyn}}{{Cantrell} \&
  {Bailyn}}{2007}]{Cantrell2007}
{Cantrell} A.~G.,  {Bailyn} C.~D.,  2007, \apj, 670, 727

\bibitem[\protect\citeauthoryear{{Cantrell}, {Bailyn}, {McClintock} \&
  {Orosz}}{{Cantrell} et~al.}{2008}]{Cantrell2008}
{Cantrell} A.~G.,  {Bailyn} C.~D.,  {McClintock} J.~E.,    {Orosz} J.~A.,
  2008, \apjl, 673, L159

\bibitem[\protect\citeauthoryear{{Cantrell} \& {et al.,}}{{Cantrell} \& {et
  al.,}}{2010}]{Cantrell2010}
{Cantrell} A.~G.,  {et al.,} 2010, \apj, 710, 1127

\bibitem[\protect\citeauthoryear{{Casares}}{{Casares}}{2007}]{Casares2007}
{Casares} J.,  2007, in {Karas} V.,  {Matt} G.,  eds, IAU Symp. Vol.~238, Black
  holes: from stars to galaxies - across the range of masses..
Cambridge University Press, pp 3--12

\bibitem[\protect\citeauthoryear{{Casares}}{{Casares}}{2010}]{Casares2010}
{Casares} J.,  2010, in J.~M.~Diego L.~J.~Goicoechea J. I. G.-S.,  Gorgas J.,
  eds, Highlights of Spanish Astrophysics V. Springer-Verlag, pp 3--14

\bibitem[\protect\citeauthoryear{{Charles} \& {Coe}}{{Charles} \&
  {Coe}}{2006}]{Charles2006}
{Charles} P.~A.,  {Coe} M.~J.,  2006, Compact Stellar X-ray Sources.
Cambridge University Press, pp 215--265

\bibitem[\protect\citeauthoryear{{Clemens}}{{Clemens}}{1985}]{Clemens1985}
{Clemens} D.~P.,  1985, \apj, 295, 422

\bibitem[\protect\citeauthoryear{{Filippenko} \& {Chornock}}{{Filippenko} \&
  {Chornock}}{2001}]{Filippenko2001}
{Filippenko} A.~V.,  {Chornock} R.,  2001, \iaucirc, 7644, 2

\bibitem[\protect\citeauthoryear{{Fryer} \& {Kalogera}}{{Fryer} \&
  {Kalogera}}{2001}]{Fryer2001}
{Fryer} C.~L.,  {Kalogera} V.,  2001, \apj, 554, 548

\bibitem[\protect\citeauthoryear{{Garnavich}, {Stanek} \&
  {Berlind}}{{Garnavich} et~al.}{1999}]{Garnavich1999}
{Garnavich} P.~M.,  {Stanek} K.~Z.,    {Berlind} P.,  1999, \iaucirc, 7276, 1

\bibitem[\protect\citeauthoryear{{Hynes}, {Charles}, {Casares}, {Haswell},
  {Zurita} \& {Shahbaz}}{{Hynes} et~al.}{2003}]{Hynes2003}
{Hynes} R.~I.,  {Charles} P.~A.,  {Casares} J.,  {Haswell} C.~A.,  {Zurita} C.,
     {Shahbaz} T.,  2003, \mnras, 340, 447

\bibitem[\protect\citeauthoryear{{Hynes}, {Haswell}, {Chaty}, {Shrader} \&
  {Cui}}{{Hynes} et~al.}{2002}]{Hynes2002}
{Hynes} R.~I.,  {Haswell} C.~A.,  {Chaty} S.,  {Shrader} C.~R.,    {Cui} W.,
  2002, \mnras, 331, 169

\bibitem[\protect\citeauthoryear{{Marsh}, {Robinson} \& {Wood}}{{Marsh}
  et~al.}{1994}]{Marsh1994}
{Marsh} T.~R.,  {Robinson} E.~L.,    {Wood} J.~H.,  1994, \mnras, 266, 137

\bibitem[\protect\citeauthoryear{{McClintock}, {Remillard}, {Heindl} \&
  {Tomsick}}{{McClintock} et~al.}{2000}]{McClintock2000a}
{McClintock} J.~E.,  {Remillard} R.~A.,  {Heindl} W.~A.,    {Tomsick} J.~A.,
  2000, \iaucirc, 7466, 1

\bibitem[\protect\citeauthoryear{{Orosz} \& {et al.}}{{Orosz} \& {et
  al.}}{2007}]{Orosz2007}
{Orosz} J.~A.,  {et al.} 2007, \nat, 449, 872

\bibitem[\protect\citeauthoryear{{{\"O}zel}, {Psaltis}, {Narayan} \&
  {McClintock}}{{{\"O}zel} et~al.}{2010}]{Ozel2010}
{{\"O}zel} F.,  {Psaltis} D.,  {Narayan} R.,    {McClintock} J.~E.,  2010,
  \apj, 725, 1918

\bibitem[\protect\citeauthoryear{{Press}}{{Press}}{2002}]{Press2002}
{Press} W.~H.,  2002, Numerical recipes in C++ : the art of scientific
  computing.
Cambridge University Press

\bibitem[\protect\citeauthoryear{{Shahbaz}, {Dhillon}, {Marsh}, {Casares},
  {Zurita} \& {Charles}}{{Shahbaz} et~al.}{2010}]{Shahbaz2010}
{Shahbaz} T.,  {Dhillon} V.~S.,  {Marsh} T.~R.,  {Casares} J.,  {Zurita} C.,
  {Charles} P.~A.,  2010, \mnras, 403, 2167

\bibitem[\protect\citeauthoryear{{Shahbaz}, {Dhillon}, {Marsh}, {Casares},
  {Zurita}, {Charles}, {Haswell} \& {Hynes}}{{Shahbaz}
  et~al.}{2005}]{Shahbaz2005}
{Shahbaz} T.,  {Dhillon} V.~S.,  {Marsh} T.~R.,  {Casares} J.,  {Zurita} C.,
  {Charles} P.~A.,  {Haswell} C.~A.,    {Hynes} R.~I.,  2005, \mnras, 362, 975

\bibitem[\protect\citeauthoryear{{Shahbaz}, {Naylor} \& {Charles}}{{Shahbaz}
  et~al.}{1993}]{Shahbaz1993}
{Shahbaz} T.,  {Naylor} T.,    {Charles} P.~A.,  1993, \mnras, 265, 655

\bibitem[\protect\citeauthoryear{{Shahbaz}, {Zurita}, {Casares}, {Dubus},
  {Charles}, {Wagner} \& {Ryan}}{{Shahbaz} et~al.}{2003}]{Shahbaz2003}
{Shahbaz} T.,  {Zurita} C.,  {Casares} J.,  {Dubus} G.,  {Charles} P.~A.,
  {Wagner} R.~M.,    {Ryan} E.,  2003, \apj, 585, 443

\bibitem[\protect\citeauthoryear{{Uemura}, {Kato}, {Pavlenko}, {Shugarov} \&
  {Mitskevich}}{{Uemura} et~al.}{1999}]{Uemura1999}
{Uemura} M.,  {Kato} T.,  {Pavlenko} E.,  {Shugarov} S.,    {Mitskevich} M.,
  1999, \iaucirc, 7303, 2

\bibitem[\protect\citeauthoryear{{Vaughan}}{{Vaughan}}{2005}]{Vaughan2005}
{Vaughan} S.,  2005, \aap, 431, 391

\bibitem[\protect\citeauthoryear{{Wood}, {Smith}, {Marshall} \& {Swank}}{{Wood}
  et~al.}{1999}]{Wood1999}
{Wood} A.,  {Smith} D.~A.,  {Marshall} F.~E.,    {Swank} J.,  1999, \iaucirc,
  7274, 1

\bibitem[\protect\citeauthoryear{{Zurita} \& {et al.,}}{{Zurita} \& {et
  al.,}}{2002}]{Zurita2002}
{Zurita} {et al.,} 2002, \mnras, 334, 999

\bibitem[\protect\citeauthoryear{{Zurita}, {Casares} \& {Shahbaz}}{{Zurita}
  et~al.}{2003}]{Zurita2003}
{Zurita} C.,  {Casares} J.,    {Shahbaz} T.,  2003, \apj, 582, 369

\end{thebibliography}


\label{lastpage}
\end{document}